\pdfoutput=1

\documentclass[prb,superscriptaddress,amsmath,amssymb,twocolumn,nofootinbib, longbibliography]{revtex4-2}
\usepackage[english]{babel}

\usepackage{graphicx,bm,amsmath}
\usepackage[usenames,dvipsnames]{xcolor}
\usepackage[colorlinks,bookmarks=false,citecolor=NavyBlue,linkcolor=Red,urlcolor=blue,
]{hyperref}

\usepackage[bbgreekl]{mathbbol}
\usepackage{xcolor}
\usepackage[normalem]{ulem}
\usepackage{algorithm}
\usepackage{braket}
\usepackage{algpseudocode}
\usepackage{comment}
\usepackage{amsthm}
\usepackage{enumerate}

\def\doi{http://dx.doi.org/}

\newcommand{\be}{\begin{equation}}
\newcommand{\ee}{\end{equation}}
\newcommand{\bec}{\begin{equation*}}
\newcommand{\eec}{\end{equation*}}
\newcommand{\bea}{\begin{eqnarray}}
\newcommand{\eea}{\end{eqnarray}}

\newcommand{\titleinfo}{Non-stabilizerness versus entanglement in matrix product states}

\newcommand{\Tr}{\text{Tr}}   
 
\begin{document}
\title{\titleinfo}
\author{M. Frau}
\affiliation{International School for Advanced Studies (SISSA), Via Bonomea 265, I-34136 Trieste, Italy}
\author{P. S. Tarabunga}
\affiliation{International School for Advanced Studies (SISSA), Via Bonomea 265, I-34136 Trieste, Italy}
\affiliation{The Abdus Salam International Centre for Theoretical Physics (ICTP), Strada Costiera 11, 34151 Trieste, Italy}
\affiliation{INFN, Sezione di Trieste, Via Valerio 2, 34127 Trieste, Italy}
\author{M. Collura}
\affiliation{International School for Advanced Studies (SISSA), Via Bonomea 265, I-34136 Trieste, Italy}
\affiliation{INFN, Sezione di Trieste, Via Valerio 2, 34127 Trieste, Italy}
\author{M. Dalmonte}
\affiliation{International School for Advanced Studies (SISSA), Via Bonomea 265, I-34136 Trieste, Italy}
\affiliation{The Abdus Salam International Centre for Theoretical Physics (ICTP), Strada Costiera 11, 34151 Trieste, Italy}
\author{E. Tirrito}
\affiliation{The Abdus Salam International Centre for Theoretical Physics (ICTP), Strada Costiera 11, 34151 Trieste, Italy}
\affiliation{Pitaevskii BEC Center, CNR-INO and Dipartimento di Fisica,
Università di Trento, Via Sommarive 14, Trento, I-38123, Italy}

\begin{abstract}
In this paper, we investigate the relationship between entanglement and non-stabilizerness (also known as magic) in matrix product states (MPSs). We study the relation between magic and the bond dimension used to approximate the ground state of a many-body system in two different contexts: full state of magic and mutual magic (the non-stabilizer analogue of mutual information, thus free of boundary effects) of  spin-1 anisotropic Heisenberg chains.   
Our results indicate that obtaining converged results for non-stabilizerness is typically considerably easier than entanglement. For full state magic at critical points and at sufficiently large volumes, we observe convergence with $1/\chi^2$, with $\chi$ being the MPS bond dimension. At small volumes, magic saturation is so quick that, within error bars, we cannot appreciate any finite-$\chi$ correction. Mutual magic also shows a fast convergence with bond dimension, whose specific functional form is however hindered by sampling errors.  
As a by-product of our study, we show how Pauli-Markov chains (originally formulated to evaluate magic) resets the state of the art in terms of computing mutual information for MPS. We illustrate this last fact by verifying the logarithmic increase of mutual information between connected partitions at critical points. By comparing mutual information and mutual magic, we observe that, for connected partitions, the latter is typically scaling much slower - if at all - with the partition size, while for disconnected partitions, both are constant in size.

\end{abstract}

\maketitle

\section{Introduction}

The past two decades have witnessed a surge in leveraging quantum information tools to tackle the long-standing challenge of the many-body problem \cite{nielsen2002quantum}. This synergy has opened avenues for novel theoretical frameworks and powerful computational techniques to describe and simulate complex quantum systems \cite{shor1999polynomial}. 
Entanglement, a key resource for performing several quantum information protocols \cite{preskill2012quantum}, has played a pivotal role in this context~\cite{plenio2014introduction,amico2008entanglement,eisert2010colloquium}. Entanglement has been discussed in the characterization of a wide range of phenomena, from real-time dynamics \cite{calabrese2005evolution}, to topological order \cite{PhysRevLett.96.110404,PhysRevLett.96.110405} and classification of states \cite{PhysRevB.83.035107,PhysRevB.84.165139}.
Moreover, a particularly useful perspective has pivoted on the role of bipartite entanglement in characterizing quantum wave functions, leading to the formulation of several classes of tensor network (TN) states \cite{Orus2014annphys, Schollwoeck2011,ran2020tensor,Banuls2023}. 

However, it is also well known that entanglement is {\it not} the only resource required for specific quantum information tasks. In particular, in the context of quantum computing, an equally essential resource is non-stabilizerness - also known as magic \cite{bravyi2005UniversalQuantumComputation,campbell2017roads}. Much like entanglement, nonstabilizerness has been quantified within the framework of resource theory using measures of nonstabilizerness.
Several measures of nonstabilizerness have been proposed in quantum information theory, with most of them relying on the notion of quasiprobability distributions \cite{wootters1987wigner,veitch2012negative,wigner1997quantum,gross2007non,hudson1974wigner}. However, in the context of many-body systems, most of these quantifiers are difficult to evaluate even numerically (see, e.g., Refs. \cite{bravyi2016improved,bravyi2016trading,howard2017robustness,heinrich2019robustness,wang2020efficiently,heimendahl2021stabilizer}), as they require in principle very complex minimization procedures. 
This computational intractability has hindered the task of quantifying nonstabilizerness beyond a few qubits, posing a major challenge in the field. 
To address this challenge, computable and practical measures of nonstabilizerness have been introduced recently: Bell magic \cite{PRXQuantum.4.010301},  stabilizer nullity \cite{Beverland2020} and  stabilizer Renyi entropies (SREs) \cite{leone2022stabilizer}. Notable
progress has been made in their experimental measurements \cite{Oliviero2022,niroula2023phase,Bluvstein2023}.

Very much like when entanglement entered the many body realm, it is presently not very clear how (and if) magic relates to physical phenomena, and how magic and entanglement are intertwined at the many-particle level. 
A very promising framework where the above question might be answered is provided by TN states. The control parameter describing their descriptive power, often referred to as bond dimension, is directly connected with the entanglement. 
Most relevantly, TN states provide presently the only theoretical framework where magic can be computed at large scale, utilizing both exact \cite{haug2023quantifying,tarabunga2024nonstabilizerness} and stochastic methods\cite{Lami2023,haug2023stabilizer,tarabunga2023manybody,tarabunga2023critical}. 

Here, we study the connection between nonstabilizerness and the bond dimension of matrix product states (MPSs) \cite{cirac2021}. We pursue a two-pronged approach. First, we investigate the broader question of how these two quantum resources, crucial for realizing states intractable to classical computers, are interrelated in many-body systems. Both nonstabilizerness and entanglement are known to be essential for such simulations. Second, we focus on the convergence properties of magic within the context of MPS simulations. Essentially, for a given variational simulation with a fixed bond dimension $\chi$, we explore how quickly these approximations converge to the true values.

We address the magic-bond dimension relationship in two distinct, yet equally intriguing, many-body scenarios: full state magic and mutual magic of ground states (GSs). These scenarios offer diverse perspectives on the interplay between magic and entanglement. In the first case, we compare a global property (magic) with correlations between arbitrary sub-system parts (entanglement). Conversely, mutual magic examines shared resources between partitions, potentially relevant to field theory due to the cancellation of all boundary terms inherent to the chosen lattice regularization.

Our investigations encompass the scaling of full-state magic for critical points in spin-1 chains, as well as within gapped phases. We observe polynomial scaling for critical points, while gapped phases exhibit saturation at low bond dimensions, where sampling errors hinder clear scaling analysis. Notably, the observed polynomial scaling is compatible with a $1/\chi^2$ dependence across all cases, suggesting a convergence rate exceeding that of bipartite entanglement. This holds for both perfect sampling and Pauli MPS representations.

Subsequently, we leverage mutual magic to study quantities independent of the UV cutoff, employing Pauli-Markov chains. The findings reveal remarkably fast convergence for this approach. Surprisingly, Pauli-Markov chains exhibit an inverse scaling of autocorrelation time, facilitating sampling at larger system sizes. However, formulating definitive statements remains challenging due to potential transition-dependent effects.

Collectively, these findings shed light on the remarkably strong connection between magic and entanglement within the context of MPS, a connection that appears robust in different classes of criticality.

The paper is structured as follows. In Sec.~\ref{Sec:SRE-MPS}, we review basic aspects of the resource theory of magic, including stabilizer Renyi entropies and their connection to mana. Moreover, we discuss the presently available methods to compute magic with MPS using both sampling-based and exact techniques. In Sec.~\ref{Sec:results}, we present our results on both full state magic, mutual magic, and mutual information for spin-1 chains. Finally, in Sec.~\ref{sec:concl}, we draw our conclusions and discuss outlooks.

\section{Nonstabilizerness in many-body system}\label{Sec:SRE-MPS}

\subsection{Resource theory for magic}

The theory of error correction based on the stabilizer formalism has motivated a resource theory of nonstabilizerness, or magic. Here, we briefly review it, and summarize the main results in addressing magic in the context of many-body theory.

We consider a system with $N$ qubit with Hilbert space $\mathcal{H}=\otimes^N_{j=1} \mathcal{H}_i$. The $N$-qubit Pauli group $\mathcal{P}_N$ encompasses all the possible Pauli strings with overall phases $\pm 1$ or $\pm i$. Mathematically, we can define $\mathcal{P}_N$ in the following way:
\be 
\mathcal{P}_N = \left \lbrace e^{i\theta \frac{\pi}{2}} \sigma_{j_1} \otimes \cdots \sigma_{j_N} | \theta, j_k=0,1,2,3   \right \rbrace .
\ee
A pure $N$-qubit state is called a stabilizer state if it satisfies certain conditions. Specifically, a stabilizer state is associated with an Abelian subgroup $\mathcal{S} \subset \mathcal{P}_N$ containing $2^N$ elements such that $S|\psi\rangle=|\psi\rangle$ for all $S\in \mathcal{S}$ (i.e.\ $S$ stabilizes $\ket{\psi}$). 
Alternatively, we can define a stabilizer state using the Clifford unitaries. The Clifford group $\mathcal{C}_N$ is defined as the normalizer of the $N$-qubit  
\be 
\mathcal{C}_N=\left \lbrace U \  \mbox{such that} \ UPU^{\dagger} \in \mathcal{P}_N \, \text{for  all} \ P \in \mathcal{P}_N \right \rbrace .
\ee
The Clifford group can be generated using the Hadamard gate, the $\pi/4$ phase gate, and CNOT gate. Stabilizer states encompass all the states that can be generated by Clifford operations acting on the computational basis state $|0\rangle^{\otimes N}$. They have a wide range of applications in quantum information \cite{PhysRevA.54.1862}, condensed matter physics \cite{KITAEV20032}, and quantum gravity \cite{pastawski2015holographic}. 

Despite of their very rich structures and large entanglement, it is well known that stabilizer states are not sufficient for quantum advantage.
Indeed, the Gottesman-Knill theorem  \cite{gottesman1997stabilizer,gottesman1998heisenberg,aaronson2004improved} gives us a constructive way to efficiently simulate Clifford circuits purely classically.  

On the other hand, universal quantum computation can be achieved through supplying magic (non-free)
states. This can be achieved by augmenting the Clifford
group with the Toffoli gate or the $\pi/8$-phase ($T$) gate, thus
unlocking the potential for universal quantum computation.

In this context, it is important to quantify the non-Clifford resources needed to prepare a given state. The amount of non-stabilizerness, or magic, of any state is measured using magic monotones. The properties required for a good monotone $\mathcal{M}$ of nonstabilizerness are: (i) $\mathcal{M} (|\psi\rangle)=0$ iff $|\psi\rangle$ is a stabilizer state (ii) nonincreasing under Clifford operations: $\mathcal{M}(\Gamma |\psi \rangle) \leq \mathcal{M}(|\psi \rangle)$, and (iii) $\mathcal{M}(|\psi\rangle \otimes |\phi \rangle) = \mathcal{M}(|\psi \rangle)+\mathcal{M}(|\phi \rangle)$ (in general, sub-additivity is sufficient). 

Previous studies primarily explored magic measures in small or weakly correlated systems, hindering our grasp of nonstabilizerness in entangled, many-body systems. This limitation stems from the exponential growth of stabilizer states and their increasingly complex geometries with increasing system size. As a consequence, calculating or numerically analyzing nonstabilizerness measures for large states becomes tremendously challenging. Only very recently this has been made possible by the developments of analytical and numerical tools to compute the SREs at large scales \cite{haug2023quantifying,tarabunga2024nonstabilizerness,Lami2023,haug2023stabilizer,tarabunga2023manybody,tarabunga2023critical, tarabunga2023magic, liu2022many, chen2023magic, passarelli2024nonstabilizerness}. 

From a quantum information perspective, understanding and measuring nonstabilizerness in MPS is crucial because it's a potential resource for achieving quantum advantage~\cite{gottesman1997stabilizer,liu2022many,gupta2024encoding}. Recent studies suggest that the computational power of a state isn't solely determined by its overall nonstabilizerness (magic density). Other characteristics, including subleading terms \cite{tarabunga2023manybody}, nonlocal components \cite{liu2022many}, and potentially topological aspects \cite{ellison2021symmetry}, also play significant roles.

Beyond its practical importance, understanding the connection between nonstabilizerness (quantum correlations) and physical phenomena holds broader significance, similar to the established importance of entanglement in this context. However, this connection remains poorly understood due to the lack of computable measures for nonstabilizerness and the methods to analyze them. Building on previous work (Refs. \cite{haug2023quantifying,Lami2023,tarabunga2023manybody,tarabunga2024nonstabilizerness,haug2023efficient,leone2024stabilizer}), we will utilize the SREs as a measure of nonstabilizerness to explore this connection within the framework of MPS.

\subsection{Stabilizer Rényi entropies}
Stabilizer Rényi Entropies (SREs) are a measure of nonstabilizerness recently introduced in Ref.~\cite{leone2022stabilizer}. For a pure quantum state $\rho$, SREs are expressed in terms of the expectation values of all Pauli strings in $\mathcal{P}_N$: 
\be \label{eq:SRE_def}
M_n \left( \rho\right)= \frac{1}{1-n} \log \left \lbrace \sum_{P \in \mathcal{P}_N} \frac{|\Tr \left(\rho P\right)|^{2n}}{d^N} \right \rbrace \  , 
\ee
with $d$ is the local dimension of the Hilbert space of $N$ qudits and $\mathcal{P}_N$ is the generalized Pauli group of $N$ qudits. The SREs have the following properties:   \cite{leone2022stabilizer} (i) faithfulness: $M_n(\rho)=0$ iff $\rho \in \text{STAB}$, (ii) stability under Clifford unitaries $C \in \mathcal{C}_N$: $M_n(C\rho C^\dagger)=M_n(\rho)$ , and (iii) additivity: $M_n(\rho_{A} \otimes\rho_{B})=M_n(\rho_{A})+M_n(\rho_{B} )$. The SREs are thus a good magic measure from the point of view of resource theory, where the free states are defined as the stabilizer states while the free operations are the Clifford unitaries. This definition is a straightforward generalization to general local dimension $d$ from the one given in Ref. \cite{leone2022stabilizer}. For $d>2$, the Pauli operators are no longer Hermitian, and thus the expectation values can be complex. For this reason, we take 
the absolute values of the expectation values $|\Tr \left(\rho P\right)|^{2n}$. 
Eq.\eqref{eq:SRE_def} can be seen as the Rényi-n
entropy of the classical probability distribution:
\be \label{eq:classicalProb}
\Xi_P= |\Tr \left(\rho P \right) |^2/d^N
\ee
It is important to note that, for spin-1/2 systems, an exception to the SRE properties mentioned above has been discussed in Ref.~\cite{haug2023quantifying} for $n<2$. An overview discussion can be found in Ref.~\cite{leone2024stabilizer}.

The definition of SREs can be extended to
mixed states by properly normalizing $\Xi_P$. For example, for
$n = 2$, the mixed state SRE is given by
\be 
\tilde{M}_2 = -\log \left( \frac{\sum_{P\in P_N} |\Tr(\rho P)|^4 }{\sum_{P\in P_N} |\Tr(\rho P)|^2} \right)
\ee
which can be seen as the Rényi-2 entropy of
\be 
\Xi_P=|\Tr(\rho P)|^2 / \sum_{P\in P_N} |\Tr(\rho P)|^2,
\ee
apart from some offset. Here, the free states are defined as
the set of mixed stabilizer states that can be obtained from pure stabilizer
states by partial tracing. This implies that, in general, SRE are not always magic monotones for mixed states, as the above set does not include more general convex combinations. Remarkably, as we discuss below, they are - under rather ubiquitous conditions - for odd-dimensional Hilbert spaces: this is the main reason why, in this work, we focus solely on spin-1 chains. 

While well motivated from the information theoretical side, full-state magic is somewhat suboptimal from a quantum many-body viewpoint, as it contains information about the edges of the system for open chains. Moreover, it has been shown that, even for simple spin models, full state magic is not necessarily associated to collective phenomena such as phase transitions~\cite{tarabunga2023critical,tarabunga2023magic}. This is similar to what happens to entanglement and Renyi entropies. 

In order to overcome such limitations, a UV cut-off independent quantity is desirable, similar to the mutual information in the context of entanglement entropies. In Ref.~\cite{tarabunga2023manybody}
, long-range magic was introduced as:
\be \label{eq:lr_magic}
L_n(\rho_{AB})= \tilde{M}_n(\rho_{AB}) - \tilde{M}_n(\rho_{B}) -\tilde{M}_n(\rho_{A}) 
\ee
where $A$ and $B$ are two separated subsystems. The long-range (or, probably better, mutual)
magic is directly reminiscent of mutual information, that has 
played a major role in characterizing the distribution of both 
classical information and quantum correlations in many-body systems. It is important to say that, similar to mutual Renyi entropies, $L_n$ will in general not being strictly positive (they will for the case $n=1/2$, as discussed below). Still, from the many-body viewpoint, they are very appealing quantities, whose connection to physical phenomena has already been pointed out. 

\begin{figure}[t!]
    \centering
    \includegraphics[width=0.47\textwidth]{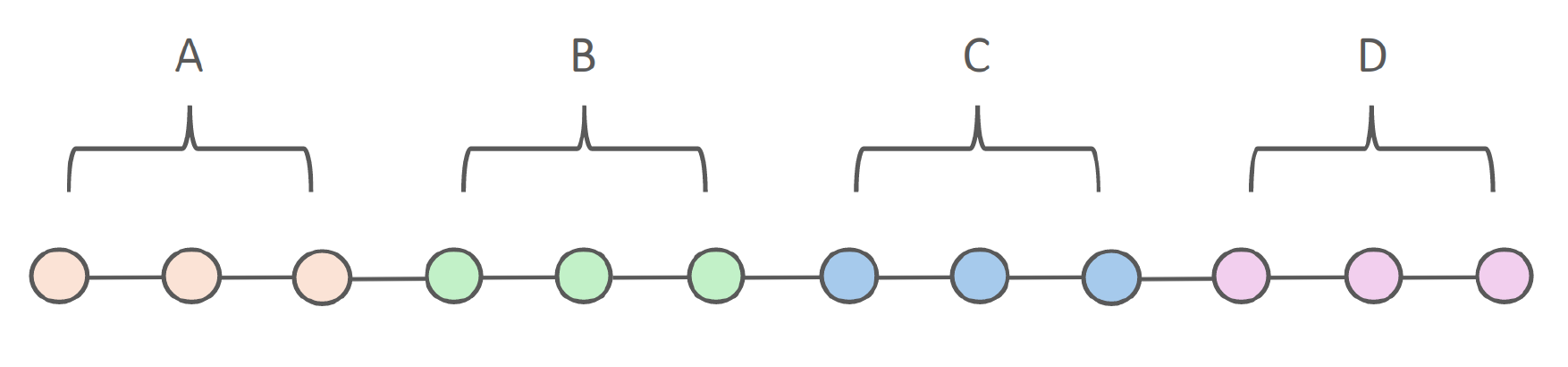}
    \caption{Schematic representation of the partitions of the chain considered for the calculation of long-range magic.}
    \label{fig:partitions}
\end{figure}

In the following, we will focus on $L_2(\rho_{AB})$, and drop the subscript for convenient notation. We will  examine two different partitioning scenarios: two connected partitions (as $B$ and $C$ in Fig. \ref{fig:partitions}); two disconnected partitions, with one positioned at the boundary and the other one in the bulk of the chain ($A$ and $C$). In all two cases, we fix the lengths of the partitions by a fixed ratio of the total number of sites in the chain, specifically $N/4$. In what follows these situations will be referred to as: case $BC$ and case $AC$.

\subsection{Connection with Mana}  \label{sec:mana}

For odd-prime-level qudits, another good measure of nonstabilizerness is the mana entropies, recently introduced in Ref. \cite{tarabunga2023critical}. To define the mana entropies for qudits with local dimension $d$, consider the generalized Pauli operators (also known as Heisenberg-Weyl operators), which are defined in terms of the shift and clock operators
    \[
    X = \sum_{k = 0}^{d-1}|k+1\rangle\langle k |, \quad Z = \sum_{k = 0}^{d-1}\omega_d^k |k \rangle \langle k |
    \]
    where $\omega_d = e^{2\pi i /d}$, as 
    \[
    T_{aa'} = \omega_d^{-\frac{a a'}{2}}Z^{a}X^{a'}
    \]
    for $a, a' \in \mathbf{Z}_d$. For a system of $N$ qudits, the Pauli strings are 
    \[
    T_{\textbf{a}}=T_{a_1a_{1'}}T_{a_2a_{2'}}\dots T_{a_Na_{N'}} .
    \]
    The phase space point operators are defined in terms of the Pauli strings as 
    \[
    A_0 = \frac{1}{d^N} \sum_\textbf{u} T_{\textbf{u}}, \quad A_{\textbf{u}} = T_{\textbf{u}}A_0T_{\textbf{u}}^\dagger
    \]
    and they provide an orthonormal basis for an operator in $\mathbf{C}^{d^N \otimes d^N}$. 
    Finally, the mana entropies are defined as
\be \label{eq:mana_def}
\mathcal{M}_n \left( \rho\right)= \frac{1}{1-n} \log \left \lbrace \sum_{\textbf{u}} \frac{|\Tr \left(\rho A_\textbf{u}\right)|^{2n}}{d^N} \right \rbrace \ . 
\ee
Mana entropies are simply SREs with the Pauli operators replaced by the phase-space point operators. The mana entropy for $n=1/2$ is equivalent to the mana \cite{veitch2012negative,veitch2014resource}, which is a genuine nonstabilizerness monotone both for pure states and mixed states.

It was shown in Ref.~\cite{tarabunga2023critical} that, whenever a state fulfils the relation
\be 
A_\textbf{u} |\psi\rangle = |\psi\rangle \;,
\ee
for a given $\textbf{u}$, mana entropies coincide with SREs for all order.

\subsection{Computability from MPS} \label{sec:mps_methods}

One of the reasons why SREs are appealing is the fact that they can often be computed much more efficiently than previously known non-stabilizerness
monotones. Even if for generic states the computational cost grows exponentially in $N$ \cite{leone2022stabilizer}, the evaluation of Pauli expectation values is, in practice, simpler than carrying out the minimization procedure involved in the definition of the robustness and min relative entropy of magic.

Moreover, for certain classes of many-body states, the computational cost
is polynomial in the number of qubits $N$. This is the case for MPS, the simplest tensor network states, that describes a pure state $|\psi \rangle$ in the following way:
\be \label{eq:mps}
|\psi \rangle=\sum_{s_1,s_2,\cdots,s_N} A^{s_1}_1 A^{s_2}_2 \cdots A^{s_N}_N |s_1,s_2,\cdots s_N \rangle
\ee
with $A_i^{s_i}$ being $\chi \times \chi$ matrices, except at the left (right)
boundary where $A_1^{s_1}$ ($A_N^{s_N}$) is a $1 \times \chi$ ($\chi \times 1$) row (column) vector. MPSs admit a useful graphical representation, where each matrix $A^s$  is interpreted as a tensor with three indices, denoted by three outer legs.

This section aims to review the three different methods to calculate the SREs: (i) Perfect Pauli sampling \cite{Lami2023,haug2023stabilizer} (ii) Pauli-Markov MPS \cite{tarabunga2023manybody} and (iii) replica Pauli-MPS \cite{tarabunga2024nonstabilizerness}.
All of these methods come with their own set of advantages and disadvantages, and thus they are applicable in different situations considered in this work. While the first two methods rely on sampling techniques, although treating the MPS ansatz exactly, suffer from accuracy dependence on the number of samples; in contrast, the latter method enables direct computation of the SREs (and other monotones) but necessitates further compression that introduces approximation, in order to keep computational cost manageable. We discuss all these elements below.

One further comment is in order: the spin chains we consider has either U(1) or SU(2) symmetry. Such symmetries are straightforwardly implemented in tensor networks. However, we opted not to include them while using sampling methods for two reasons: (i) this allows us to make general statements; (ii) the sampling procedure can be formulated in terms of few-site operations only: while this makes it not more efficient (including symmetries is straightforward), it facilitates comparisons between different paramenter regimes; and (iii) for most cases, full state magic will converge so quickly with bond dimension that including symmetries would have made finite-bond dimension corrections hard to characterize.

\subsubsection{Perfect Pauli sampling for MPS} 
This method is based on Tensor Networks perfect sampling algorithms \cite{Stoudenmire2010, Ferris2012} and on the fact that the SREs can be evaluated as 
\be \label{eq:estimator_n}
M_n = \frac{1}{1-n} \log  \left\langle \Xi_{P}^{n-1} \right\rangle_{\Xi_P} -N\log d
\ee
for $n\neq 1$, and
\be \label{eq:estimator_1}
M_1 =   -\left\langle\log \Xi_P \right\rangle_{\Xi_P} - N\log d
\ee
for $n=1$, where $\langle ... \rangle_{\Xi_P}$ is the average over the probability distribution $\Xi_P$ in Eq.~(\ref{eq:classicalProb}). The crucial ingredient here is the capability to extract a Pauli string $P_{\boldsymbol{\alpha}}=\sigma^{\alpha_1}_1 \cdots \sigma^{\alpha_N}_N$, for a specific index configuration $\boldsymbol{\alpha}=(\alpha_1, \alpha_2, \cdots, \alpha_N)$, with the correct probability. The key observation relies on the fact that any probability can be rewritten as
\be \label{eq:condprob} 
\Xi_{P_{\boldsymbol{\alpha}}} = p(\alpha_1) p(\alpha_2|\alpha_1) \cdots p(\alpha_N|\alpha_1 \alpha_2\dots \alpha_{N-1}),
\ee
where $p(\alpha_j|\alpha_1 \alpha_2 \cdots \alpha_{j-1})$ is the conditional probability to get a Pauli matrix $\sigma^{\alpha_j}$ at site $j$, given the partial configuration $\sigma_{1}^{\alpha_1}\cdots \sigma_{j-1}^{\alpha_{j-1}}$ and marginalising over the remaining lattice sites. 
By iterating over this chain relation, each conditional probability can be efficiently computed for a MPS.
In particular, after a complete sweep along the entire system, and a total computational cost $O(N\chi^3)$, we generate strings of Pauli operators
$P_{\boldsymbol{\alpha}}$ associated exactly with the correct probability $\Xi_{P_{\boldsymbol{\alpha}}}$~\cite{Lami2023,haug2023stabilizer}. 

The value of $M_n$ (or $M_1$) can be estimated using Eq.s~(\ref{eq:estimator_n},\ref{eq:estimator_1}) averaging over a finite number of samples $\mathcal{S}$; the accuracy of the method of course increases with the number of samples. Indeed, the difference between the average value of $M_n$ obtained by sampling the distribution $\Xi_P$ and 
the actual value $M_n$ is of the order of $O(1/\sqrt{S})$. In addition, in the typical cases where the SREs is extensive, the number of samples required to estimate the SREs within a given error scales polynomially with $N$ for $n=1$, and it is exponential in $N$ for $n \neq 1$. \cite{Lami2023,haug2023stabilizer,tarabunga2023manybody}.

It is worth mentioning two additional methodological aspects. Perfect Pauli sampling for MPS can be enhanced by employing a clever biasing strategy, enabling direct access to the stabilizer nullity~\cite{lami2024learning}, which can be seen as well as the $n\to\infty$ limits of the SREs~\cite{tarabunga2024nonstabilizerness}. Additionally, improved sampling strategies can also be applied to SRE, as demonstrated in Ref.~\cite{ballarin2024optimal}.

\subsubsection{Pauli-Markov MPS}
The Pauli-Markov method aims to estimate the SREs by a direct sampling of the Pauli strings using Monte Carlo scheme \cite{tarabunga2023manybody}. Unlike perfect Pauli sampling, any probability distribution $\Pi_P$ which only depends explicitly on the expectation value of $P$ can be sampled. It is possible to exploit this flexibility to implement various tricks to reduce the statistical errors (see Refs. \cite{tarabunga2023manybody, tarabunga2023critical} for more details). 

One important application of Pauli-Markov is to estimate the long-range magic. To do this, we first rewrite Eq. \eqref{eq:lr_magic} as follows:
\begin{equation}
    L(\rho_{AB}) = I(\rho_{AB}) - W(\rho_{AB}),
\end{equation}
where 
\begin{equation}
\resizebox{.89\hsize}{!}{$W(\rho_{AB}) = -\log \left( \frac{\sum_{P_A \in \mathcal{P}_A} |\Tr(\rho_A P_A)|^4 \sum_{P \in \mathcal{P}_B} |\Tr(\rho_B P_B)|^4}{\sum_{P_{AB} \in \mathcal{P}_{AB}}  |\Tr(\rho_{AB} P_{AB})|^4} \right)$},
\end{equation}
and $I(\rho_{AB})=S_2(\rho_{A}) + S_2(\rho_{B}) - S_2(\rho_{AB})$ is the Rényi-2 mutual information. If one is to sample according to $\Pi_{P_{AB}} \propto \Tr(\rho_{AB} P_{AB})^4$, we can estimate $W(\rho_{AB})$ by
\begin{equation} \label{eq:estimator_w}
    W(\rho_{AB}) =  -\log  \left\langle \frac{ |\Tr  (\rho_A P_A)|^4 |\Tr  (\rho_B P_B)|^4 }{|\Tr  (\rho_{AB} P_{AB})|^4}\right\rangle_{\Pi_{P_{AB}}}, 
\end{equation}
where $P_{AB}$ is decomposed as $P_{AB}=P_A \otimes P_B$. Similarly, the Rényi-2 mutual information can be estimated by
\begin{equation} \label{eq:estimator_i2}
I(\rho_{AB}) =  -\log  \left\langle \frac{| \Tr  (\rho_A P_A)|^2 |\Tr  (\rho_B P_B)|^2 }{|\Tr  (\rho_{AB} P_{AB})|^2}\right\rangle_{\Xi_{P_{AB}}}.
\end{equation}

Pauli-Markov method is applicable with MPS since any expectation values are efficiently computable. For a single-site update scheme, if the candidate site is chosen randomly with uniform probability, the distance between the current site and the modified site is on average linear in $N$. Therefore, the cost for each update in MPS is $O(N \chi^3)$. Note that randomly choosing the modified site, as opposed to sweeping procedure, is important for ensuring detailed balance, as well as reducing the autocorrelation time.

\subsubsection{Replica Pauli-MPS}
Since the SREs are expressed in terms of Pauli expectation values, it would be advantageous to store the expectation values in an efficient way. This can be achieved simply by representing the state in the Pauli basis. If a state is efficiently represented by an MPS with bond dimenion $\chi$, its exact representation in the Pauli basis can also be written as an MPS with bond dimension $\chi^2$, i.e., the Pauli-MPS \cite{tarabunga2024nonstabilizerness}. In this formalism, the SRE of index $n$ can be evaluated as the contraction of $2n$ replicas of Pauli-MPS.

Compared to the original replica trick formulation \cite{haug2023quantifying}, the advantage of replica Pauli-MPS method comes in two aspects. First, the physical dimension of the intermediate MPS to compute the SRE of index $n$ is constantly $d^2$ using the replica Pauli-MPS approach, while Ref. \cite{haug2023quantifying} requires a physical dimension of $d^{2(n-1)}$, which grows exponentially with $n$. Since the bond dimension is $\chi^{2n}$ in both methods, the cost of exact contraction to compute the SRE is $O(N d^2 \chi^{6n})$ and $O(N d^{2(n-1)} \chi^{6n})$, respectively. Second, since the intermediate MPS in the replica Pauli-MPS method is obtained by repeated MPO-MPS multiplication, one can sequentially compress the resulting MPS after every iteration using standard tensor network routines. Although this is at the cost of computing an approximation to the SRE, the error of the truncation can be monitored, e.g., by doing standard convergence analysis. From the practical point of view, this controlled approximation scheme is crucial to be able to apply the method beyond the very small bond dimension that can be afforded using exact contraction. Assuming all intermediate MPS has bond dimensions of order $\chi$, the computational cost of approximate contraction is $O(N d^2 \chi^4)$, that significantly improves on scaling with bond dimension compared to the exact contraction.

Further, accessing the SRE of any partition is very simple in the Pauli-MPS formalism. Indeed, any site can be traced out simply by projecting the site to identity. Then, the SRE of a given partition can be obtained in a similar way as above by considering the Pauli-MPS after tracing out the sites outside of the partition.

\begin{figure}[t!]
        \centering
        \includegraphics[width=0.47\textwidth]{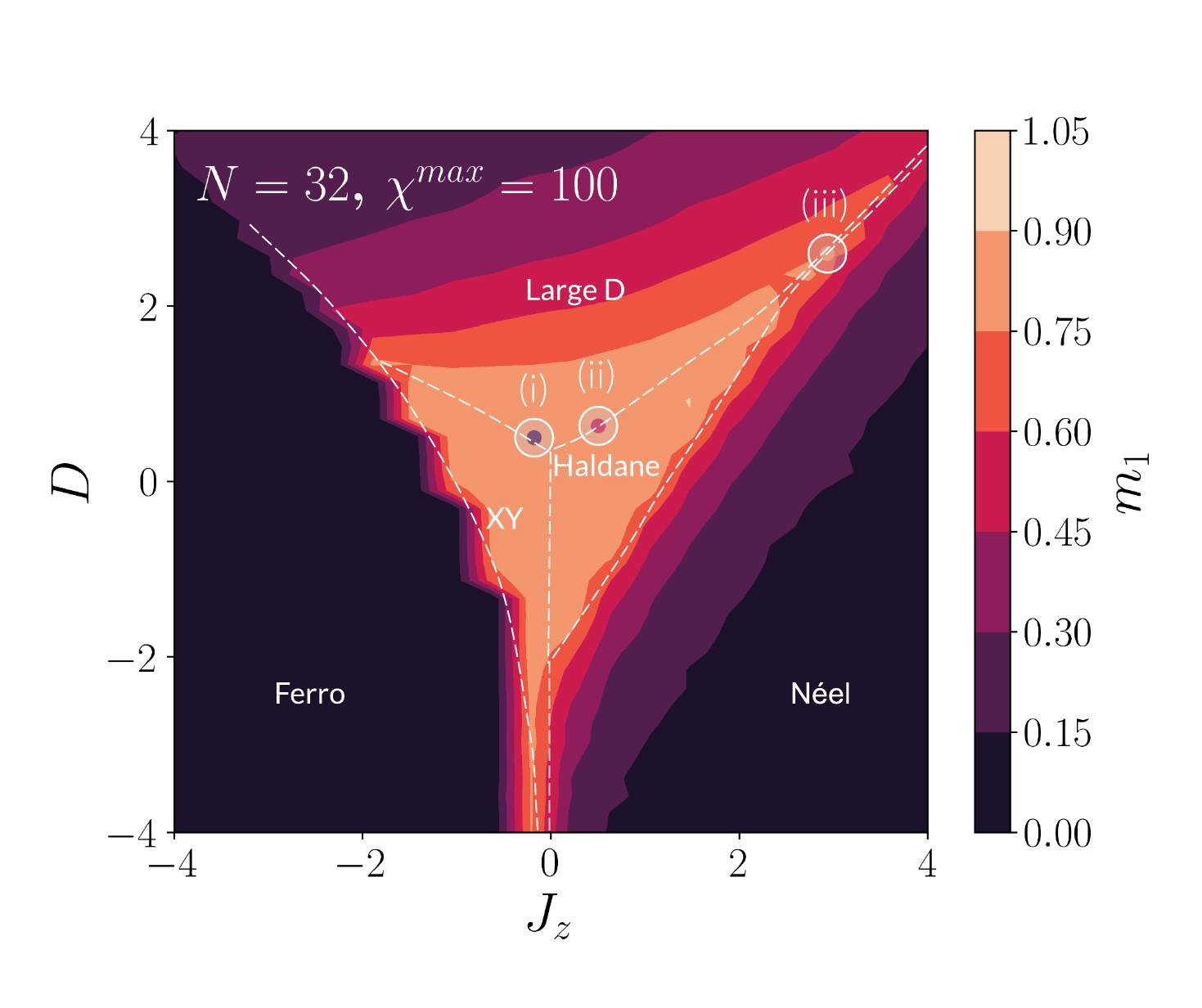}
        \caption{Phase diagram of the $S = 1$ $XXZ$ chain with uniaxial single ion-type anisotropy. The three marked points correspond to the transitions examined in the rest of the paper: (i) Large D-XY ($J_z \sim -0.183$, $D \sim 0.5$); (ii) Haldane-Large D ($J_z \sim 0.5$, $D \sim 0.635$) and (iii) Haldane-Néel ($J_z \sim 2.93$, $D \sim 2.6$). $m_1$ is computed with perfect Pauli sampling with $N_S = 10^3$ samples. }
        \label{fig:phasediagram}
\end{figure}

\section{SRE in spin-1 XXZ chain}\label{Sec:results}
We consider a $S = 1$ $XXZ$ chain with uniaxial single ion-type anisotropy:
\be \label{eq:HeisHamiltonian}
H = \sum_{i =1}^N[S^x_iS^x_{i+1}+S^y_iS^y_{i+1}+J_zS^z_iS^z_{i+1}]+D\sum_{i = 1}^N S^{z2}_i
\ee
where $S^{\alpha}$'s, $\alpha = x, y, z$, are the spin-1 operators, $J_z$ is the easy-axis anisotropy, and $D$ is the single-ion anisotropy. The model has a global $U(1)$ symmetry corresponding to the conservation of total magnetization $\sum_i S^z_i$, and here we consider the ground states at the sector of zero total magnetization. The model displays a rich phase diagram, as sketched in Fig. \ref{fig:phasediagram}, making it a good playground to explore magic in phase transitions \cite{PhysRevB.67.104401}.

For $J_z > 0$, the model hosts three phases (with increasing D): the antiferromagnetic Néel order, the symmetry-protected topological (SPT) Haldane phase, and the large-D trivial phase. The Néel to Haldane transition is an Ising transition, while the Haldane to large-D transition is a Gaussian transition.
    
For $Jz \leq 0$ the model hosts a ferromagnetic phase, the large-D trivial phase and two different XY phases: for large negative $D$ (XY2) $\langle S^x_i S^x_j \rangle$ and $\langle S^x_i S^x_j \rangle$ decay exponentially, while for small negative $D$ (XY1) they decay with a power law, therefore they can be regarded as two different phases. The XY to Haldane and XY to large-D transitions are BKT transitions, while XY to ferromagnetic and large-D to ferromagnetic are first order transitions. 

In what follows, we consider three different transitions that are present in the model's phase diagram: Haldane-Néel (Ising transition at $J_z \sim 2.93$, $D \sim 2.6$), Haldane-LargeD (Gaussian transition at $J_z \sim 0.5$, $D \sim 0.635$) and Large D-XY (BKT transition at $J_z \sim -0.183$, $D \sim 0.5$) \cite{PhysRevB.67.104401}.
\begin{figure}[t!]
        \centering
        \includegraphics[width=0.47\textwidth]{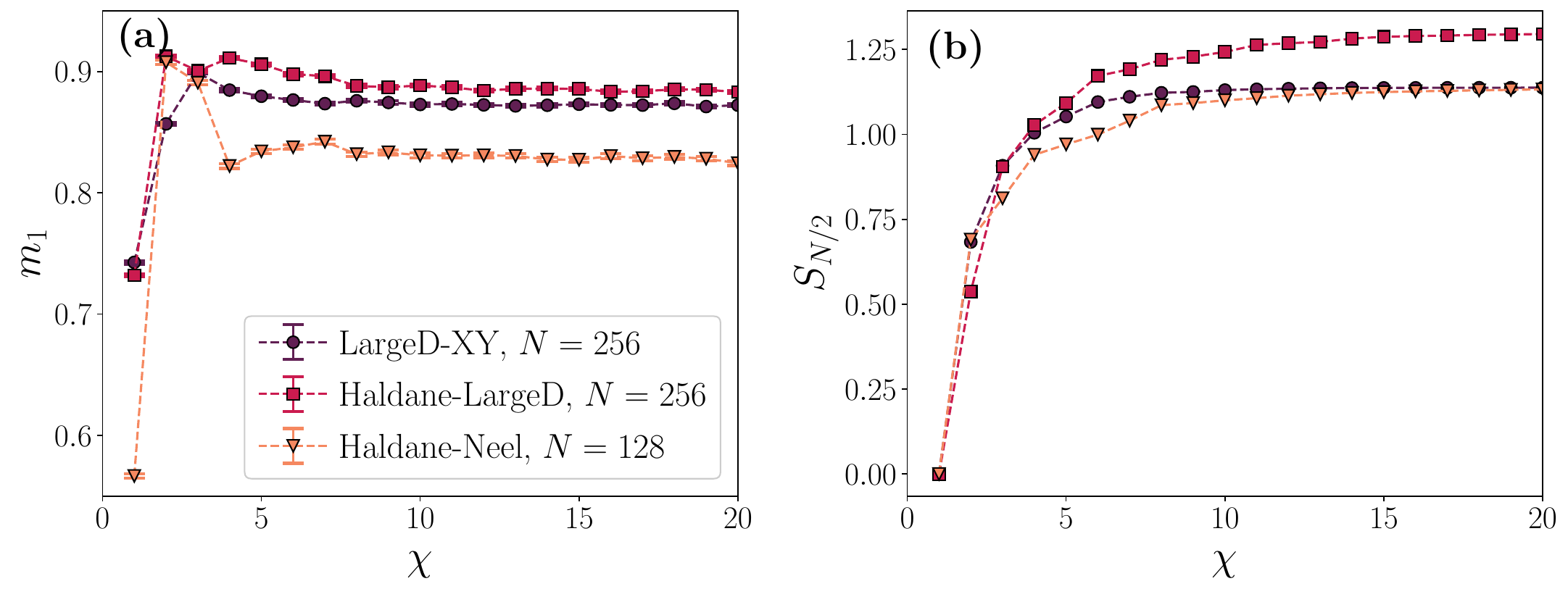}
        \caption{\textbf{(a)} The SRE density $m_1$ computed via perfect Pauli sampling for various bond dimensions $\chi$. The number of samples is $N_S = 10^4$. \textbf{(b)} The entanglement entropy $S_{N/2}$ for various bond dimensions $\chi$.}
        \label{fig:m1andsent}
\end{figure}

The SRE in $S = 1$ $XXZ$ chain was recently investigated in Ref. \cite{tarabunga2023manybody} using Pauli-Markov method. It was shown that, while the full-state magic appears rather featureless at the critical points, long-range magic is able to identify the transitions.

To obtain the MPS approximation of the ground state with a given bond dimension $\chi$, we perform DMRG simulations using the iTensor package \cite{ITensor,ITensor-r0.3}. We then employ the methods outlined in Sec. \ref{sec:mps_methods} to compute the SRE. The full-state magic is generally linear in the system size $N$, and thus we will consider the SRE density $m_n=M_n/N$. Note that, since the phase-space operator $A_\textbf{0}$ commutes with the Hamiltonian of the $S=1$ XXZ chain, the ground state SREs coincide with the mana entropies (see Sec. \ref{sec:mana}). 

For the Pauli-Markov method, due to the presence of $U(1)$ symmetry, a two-site update scheme is required to sample only the Pauli strings that are compatible with the symmetry. To this end, we generate the candidate Pauli string  $P'$ by randomly multiplying the current Pauli string $P$ with either $Z_i$ or $X_i^\dagger X_j$. Moreover, we set the probability to multiply with $Z_i$ or $Z_i^\dagger$ to be equal, so that detailed balance is satisfied.

Fig. \ref{fig:phasediagram}, obtained utilizing perfect Pauli sampling, illustrates the variation of $m_1$ within the model's phase diagram. Notably, magic reaches its peak within the topological phases (Haldane and XY) and rapidly declines in the trivial phases (Ferromagnetic, Large D, and Néel). However, no distinct characteristics are observed at the critical point, consistently with previous findings \cite{tarabunga2023manybody}.

\begin{figure}[t!]
    \centering
    \includegraphics[width=0.47\textwidth]{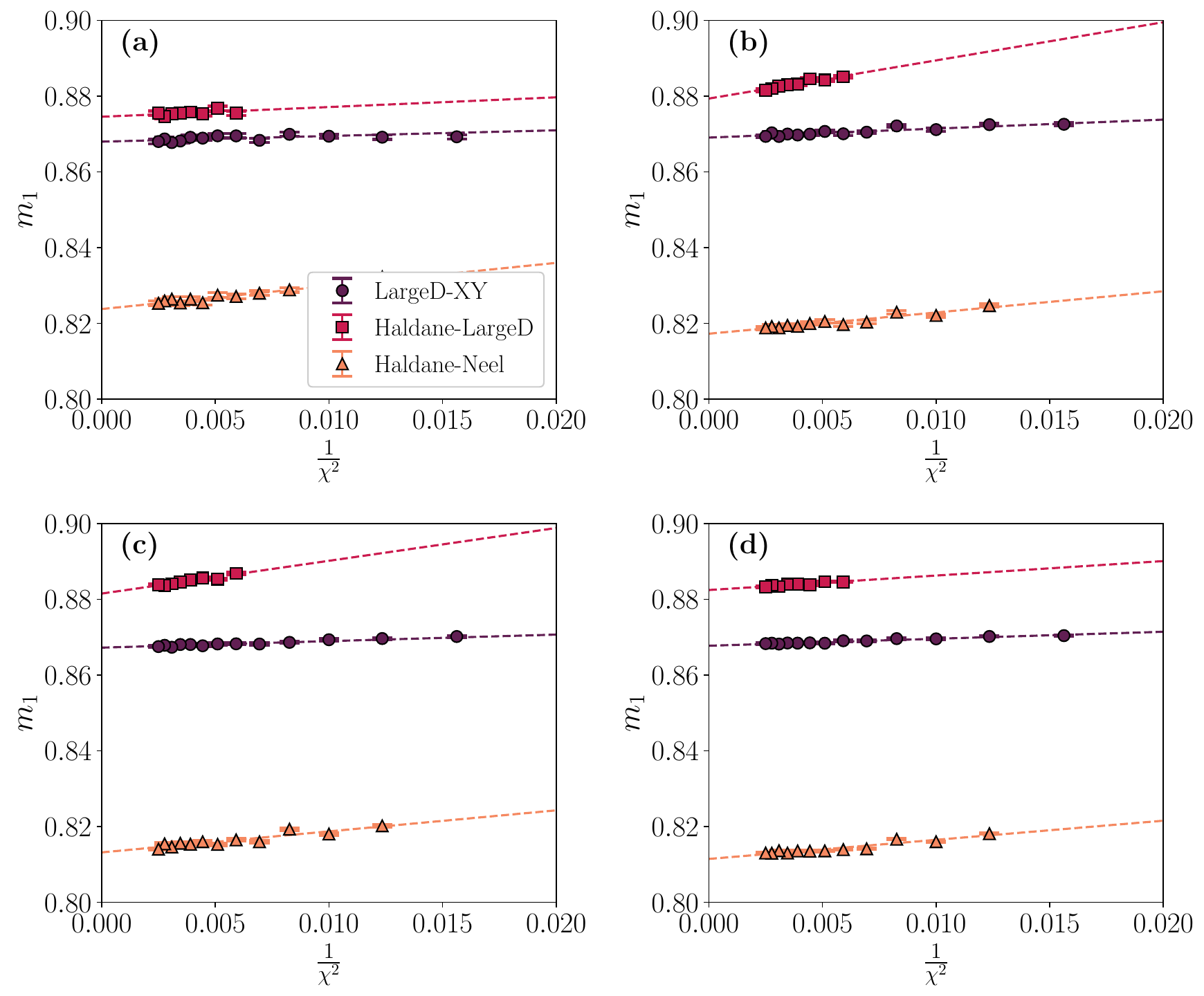}
    \caption{Scaling of the SRE density $m_1$ with the bond dimension $\chi$ in different critical points. The number of samples is $N_S = 10^4$. The different panels correspond to different system sizes: \textbf{(a)} $N = 128$; \textbf{(b)} $N = 256$; \textbf{(c)} $N = 512$ and \textbf{(d)} $N = 1024$.}
    \label{fig:m1scaling}
\end{figure}
\begin{figure}[t]
    \centering
    \includegraphics[width=0.47\textwidth]{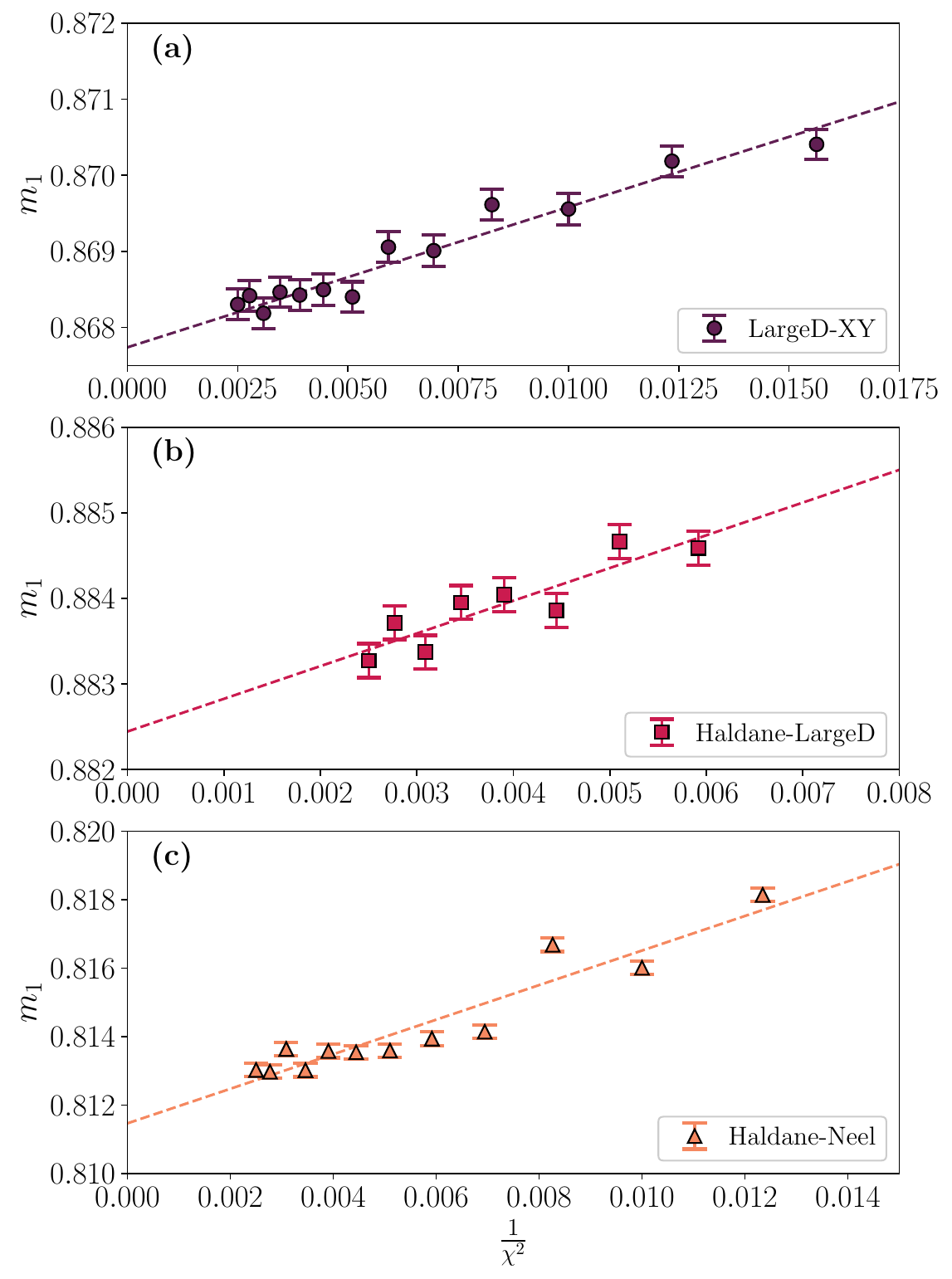}
    \caption{Scaling of the SRE density $m_1$ with the bond dimension $\chi$ in critical points at fixed $N = 1024$. The number of samples is $N_S = 10^4$ The different panels correspond to different transitions: \textbf{(a)} Large D-XY transition; \textbf{(b)} Haldane-Large D transition and \textbf{(c)} Haldane-Néel transiton.}
    \label{fig:m1scaling_1024}
\end{figure}

\subsection{Scaling of full-state magic}
We calculate the SRE and study how it scales with the bond dimension of the MPS. Initially, we utilize perfect Pauli sampling, focusing on the $n = 1$ SRE, where the number of samples only needs to scale polynomially with system size to achieve a desired level of accuracy in estimation. In Fig. \ref{fig:m1andsent}, the SRE density $m_1$ is compared with the entanglement entropy at half chain, $S_{N/2}$. The graph displays three distinct transitions: Haldane-Néel, Haldane-Large D, and Large D-XY.  We observe that convergence of non-stabilizerness occurs at smaller bond dimensions compared to those required for entanglement entropy. Indeed, as shown in Fig. \ref{fig:m1andsent}a, a bond dimension of $\chi \sim 8$ is sufficient for an accurate computation of magic density, presenting a significant practical benefit.

Moreover, in Fig. \ref{fig:m1scaling} and Fig. \ref{fig:m1scaling_1024}, we demonstrate that $m_1$ scales as $\frac{1}{\chi^2}$ at critical points. Fig. \ref{fig:m1scaling} shows the linear dependence of $m_1$ from $\frac{1}{\chi^2}$ for different system sizes ($N \in \{128,256,512,1024 \}$), while Fig. \ref{fig:m1scaling_1024} provides a detailed view of the largest system size. The dashed lines in these figures represent the fits performed with the functional form $m_1(1/\chi^2) = m_0 + c~ 1/\chi^2 $. Note that the slope $c$ is size independent in the Large D-XY and Haldane-Neel transitions, while it changes non-monotonically with $N$ in the Haldane-Large D case. Regarding the intercepts, we expect them to be size invariant since the SRE is generally linear with size $N$ and what we're computing is its density, namely $m_1 = M_1/N$. Indeed, this expectation holds true for the Large D-XY and Haldane-Large D transitions but not for the Haldane-Néel one, which appears to exhibit a more pronounced finite-size effect.

\begin{figure}[t!]
        \centering
        \includegraphics[width=0.5\textwidth]{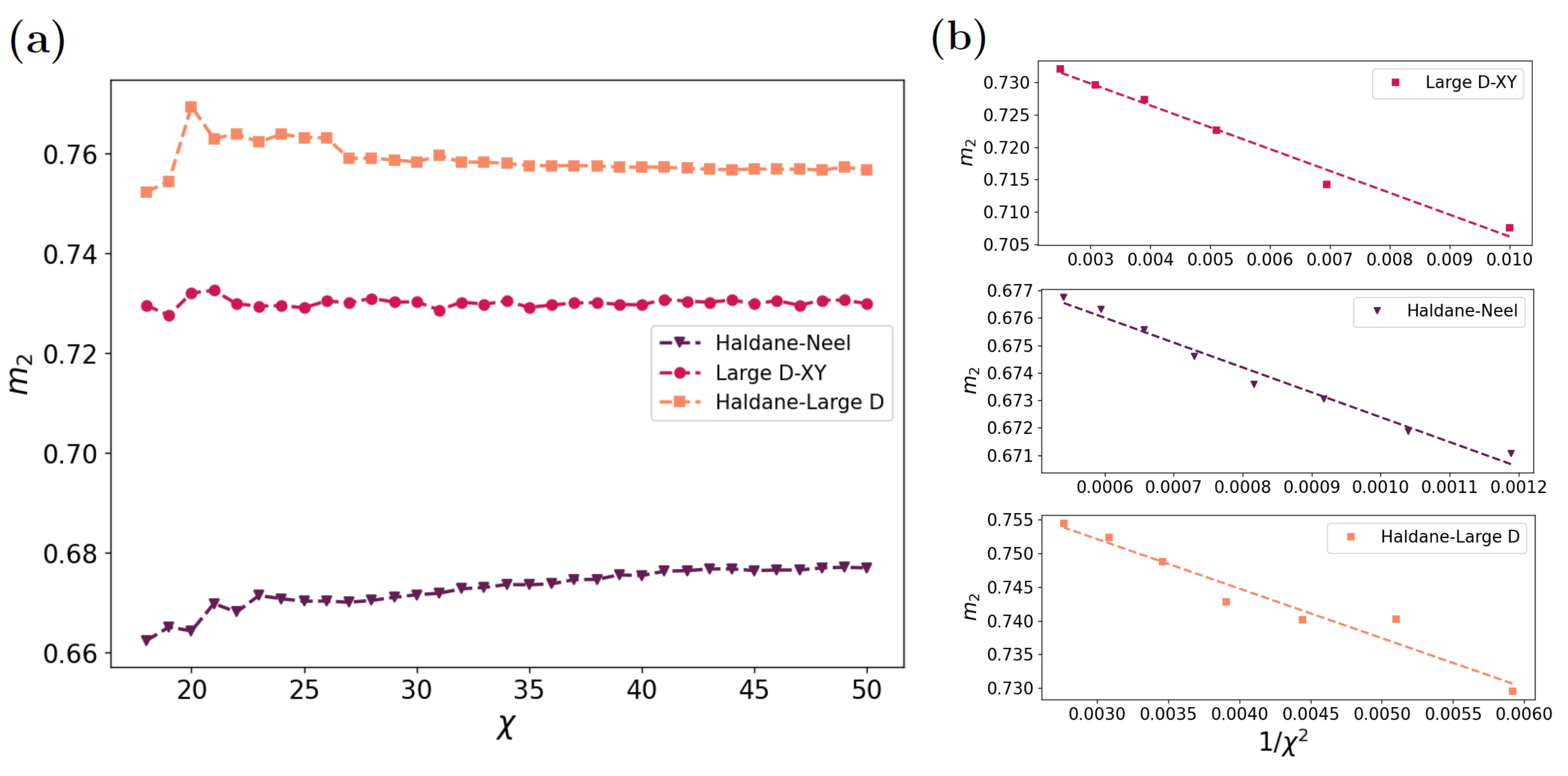}
        \caption{\textbf{(a)} The SRE density $m_2$ computed via Pauli-MPS for various bond dimensions $\chi$.  \textbf{(b)} Scaling of the SRE density $m_2$ with the bond dimension $\chi$ in the critical points at fixed $N=64$. }
        \label{fig:m2vschi}
\end{figure}

Moreover, in Fig. \ref{fig:m2vschi}, we show the behavior of $m_2$ as a function of bond dimension $\chi$ obtained from Pauli-MPS with bond dimension $ \chi_P=2*\chi$. In Fig. \ref{fig:m2vschi}a we show $m_2$ vs $\chi$ in the three distinct transitions: Haldane-Neel, Haldane-Large F and Large D-XY. We observed that the convergence of the $m_2$ occurs at small bond dimension. 
In Fig  \ref{fig:m2vschi}b,  we show that $m_2$ scales as $\frac{1}{\chi^2}$ at the critical points for system size $N=64$. In the three different panels, the dashed lines represent the fits performed with the functional form $m_2(1/\chi^2)=m_0 + c~ 1/ \chi^2$.

\subsection{Mutual information and long-range magic scaling}

To evaluate long-range magic we employ the Pauli-Markov sampling technique. This choice is motivated by the limitations encountered with perfect Pauli sampling when estimating the quantities defined in Eq.s~\eqref{eq:estimator_w}, \eqref{eq:estimator_i2}. Perfect Pauli is efficient in these two scenarios: when $\rho$ is a pure state; when $\rho$ is the reduced density matrix of a partition that falls at the boundary of a larger pure state (both rightmost subsystem, in that case one should sample with the right-normalized form of the MPS, and leftmost, where instead one should use the left-normalized form). However, it is not efficient in the case of two (possibly disconnected) partitions.

\begin{figure}[t!]
    \centering
    \includegraphics[width=0.47\textwidth]{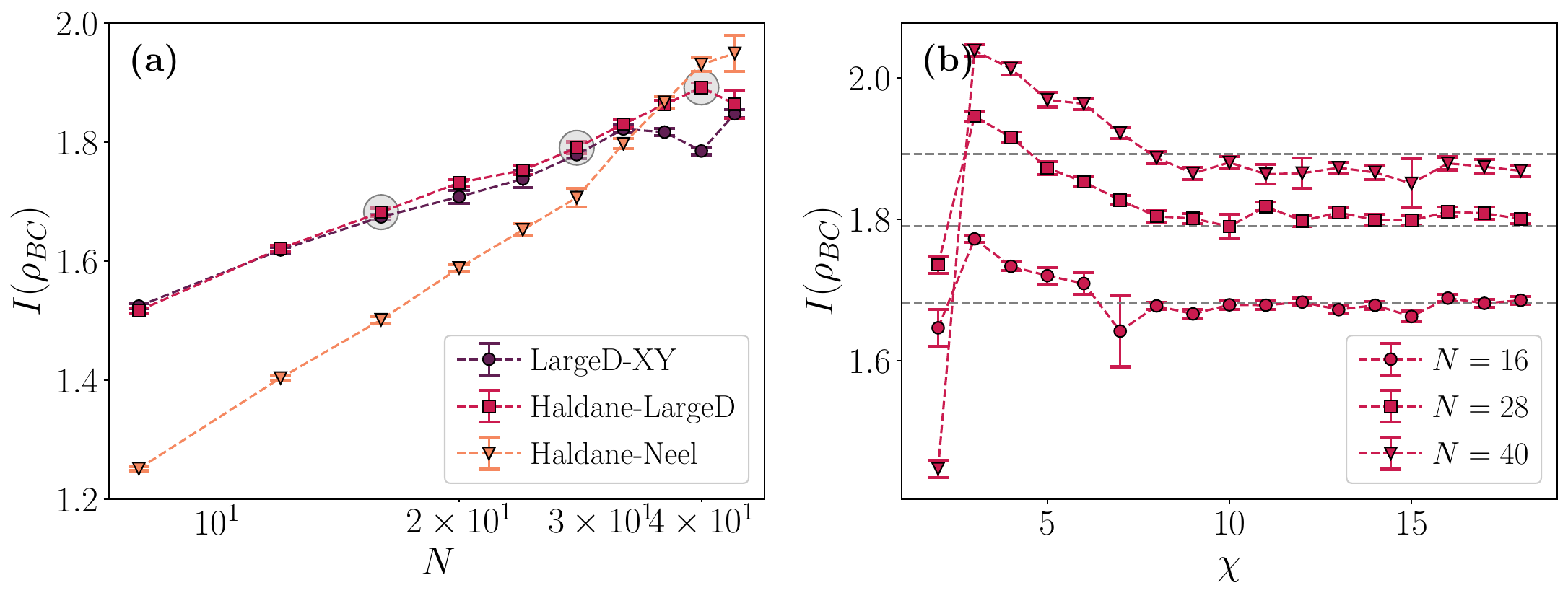}
    \caption{Mutual information scalings for connected partitions (case $BC$). The number of samples is $N_S = 10^6$. Panel \textbf{(a)}: scaling with the size for different transitions (in log scale). The three circles correspond to the sizes used for the scaling in the bond dimension $\chi$. The bond dimension here is fixed at $\chi = 20$. Panel \textbf{(b)}: scaling with the bond dimension $\chi$ in the Haldane-Large D transition (Gaussian transition at $J_z \sim 0.5$, $D \sim 0.635$), for different sizes. The dashed lines here correspond the value of the mutual information at $\chi = 20$.}
    \label{fig:mutualIscaling_connected}
\end{figure}

\begin{figure}[t!]
    \centering
    \includegraphics[width=0.47\textwidth]{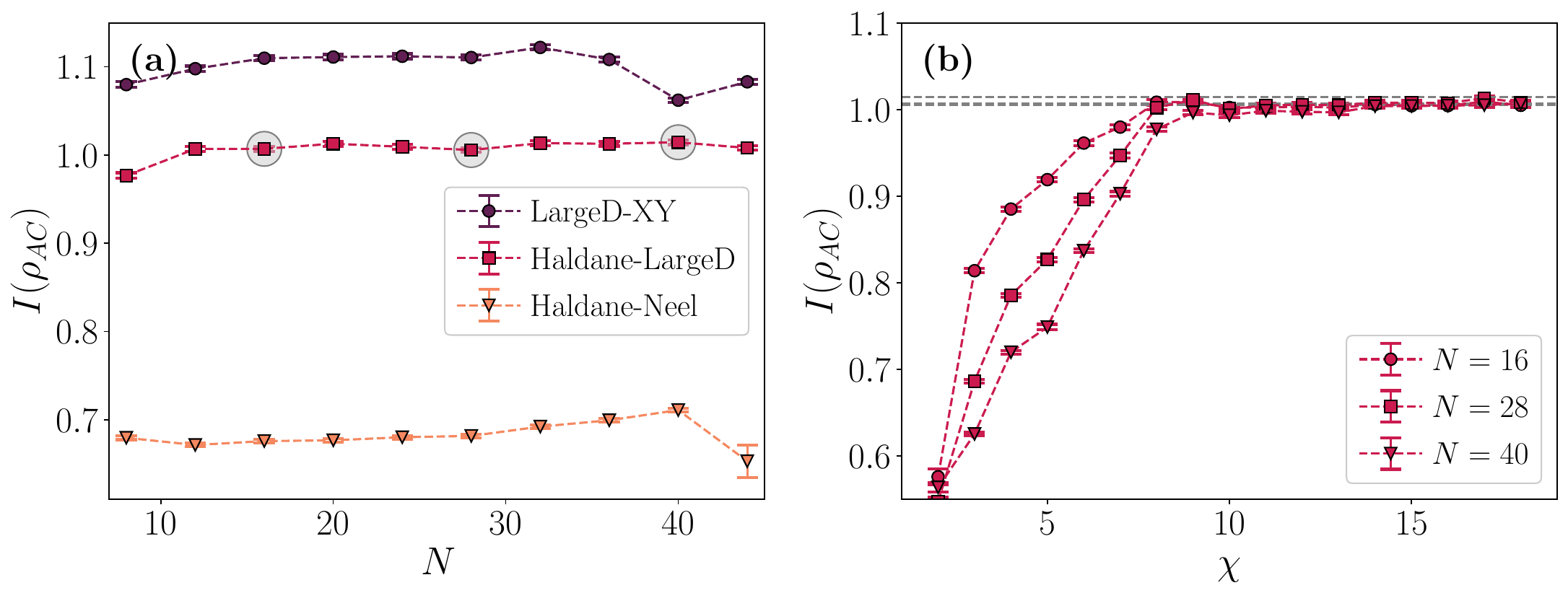}
    \caption{Mutual information scalings for disconnected partitions (case $AC$). The number of samples is $N_S = 10^6$. Panel \textbf{(a)}: scaling with the size for different transitions. The three circles correspond to the sizes used for the scaling in the bond dimension $\chi$. The bond dimension here is fixed at $\chi = 20$. Panel \textbf{(b)}: scaling with the bond dimension $\chi$ in the Haldane-Large D transition (Gaussian transition at $J_z \sim 0.5$, $D \sim 0.635$), for different sizes. The dashed lines here correspond the value of the mutual information at $\chi = 20$.}
    \label{fig:mutualIscaling_disconnected}
\end{figure}

\begin{figure}[t!]
    \centering
    \includegraphics[width=0.47\textwidth]{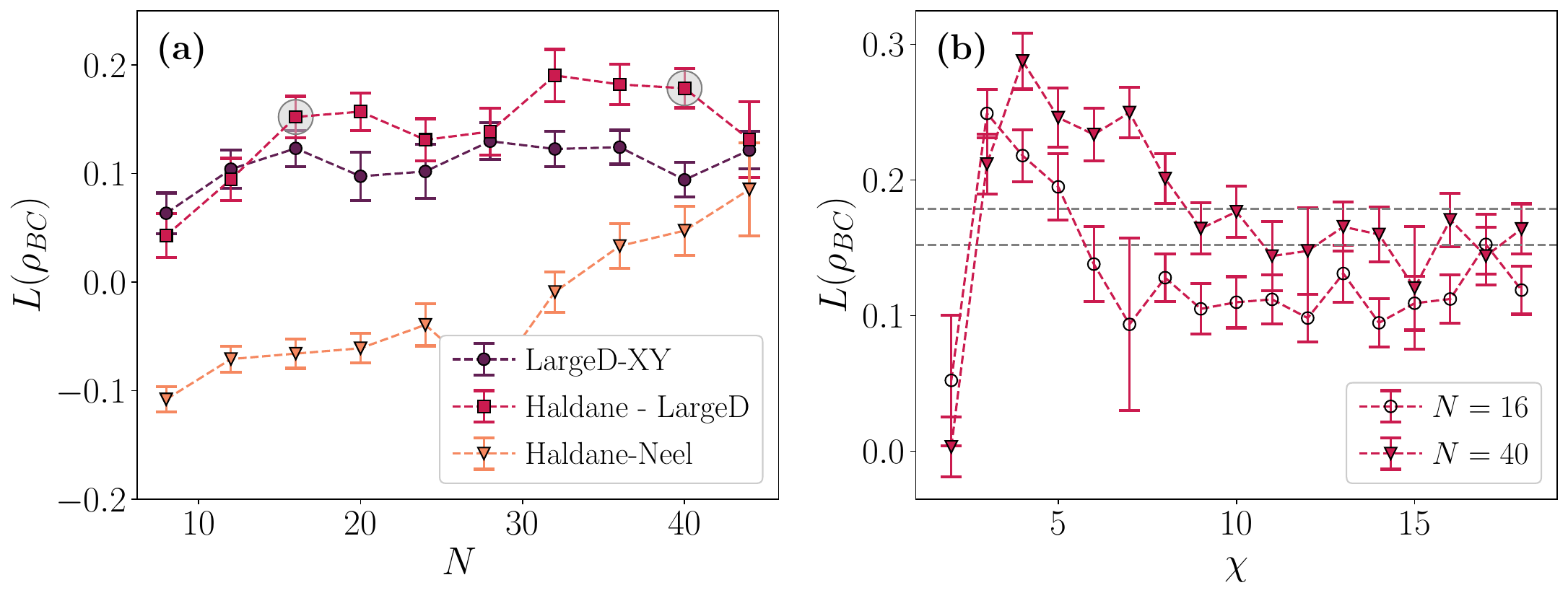}
    \caption{Long-range magic scalings for connected partitions (case $BC$). The number of samples is $N_S = 10^6$. Panel \textbf{(a)}: scaling with the size for different transitions. The two circles correspond to the sizes used for the scaling in the bond dimension $\chi$. The bond dimension here is fixed at $\chi = 20$. Panel \textbf{(b)}: scaling with the bond dimension $\chi$ in the Haldane-Large D transition (Gaussian transition at $J_z \sim 0.5$, $D \sim 0.635$), for different sizes. The dashed lines correspond the value of the long-range magic at $\chi = 20$.}
    \label{fig:LRMscaling_connected}
\end{figure}

\begin{figure}[t!]
    \centering
    \includegraphics[width=0.47\textwidth]{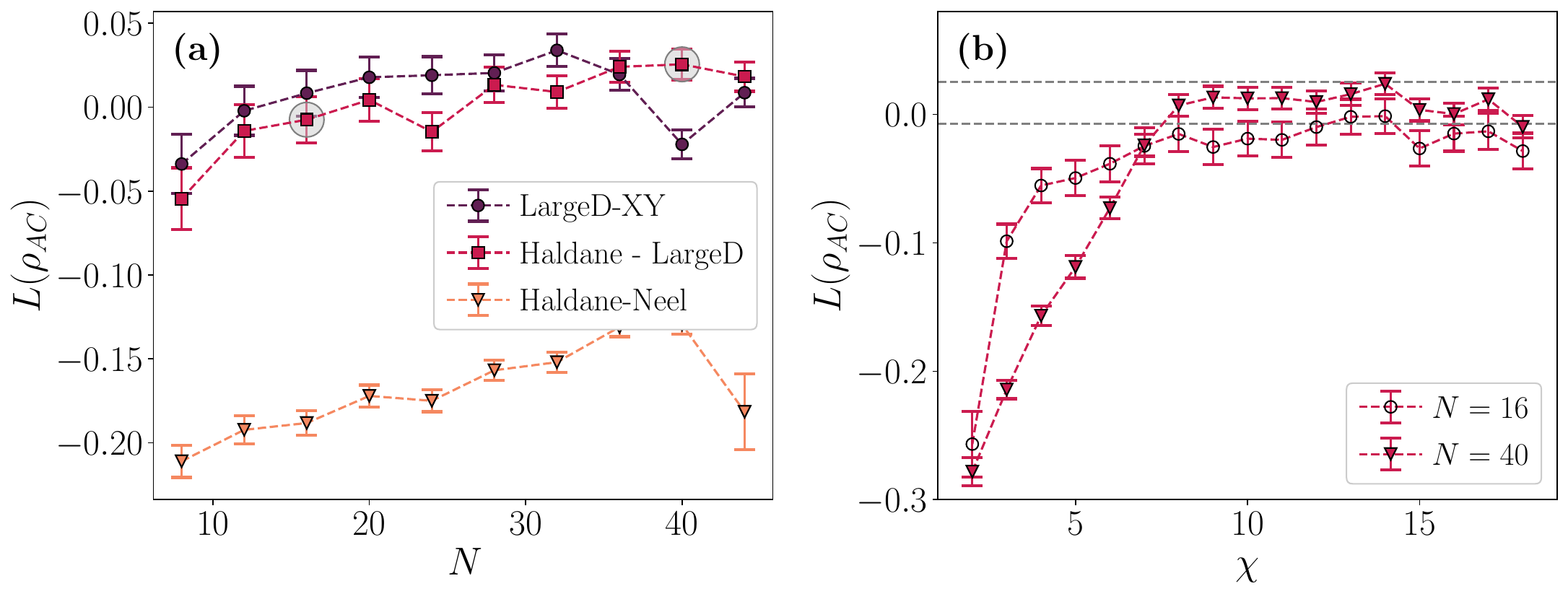}
    \caption{Long-range magic scalings for disconnected partitions (case $AC$). The number of samples is $N_S = 10^6$. Panel \textbf{(a)}: scaling with the size for different transitions. The two circles correspond to the sizes used for the scaling in the bond dimension $\chi$. The bond dimension here is fixed at $\chi = 20$. Panel \textbf{(b)}: scaling with the bond dimension $\chi$ in the Haldane-Large D transition (Gaussian transition at $J_z \sim 0.5$, $D \sim 0.635$), for different sizes. The dashed lines correspond the value of the long-range magic at $\chi = 20$.}
    \label{fig:LRMscaling_disconnected}
\end{figure}

We first focus on the mutual information $I$, obtained by means of the estimator defined in Eq. \eqref{eq:estimator_i2}. Results are presented in Fig. \ref{fig:mutualIscaling_connected} and Fig. \ref{fig:mutualIscaling_disconnected}, for case $BC$ and $AC$ respectevely. In particular, Fig. \ref{fig:mutualIscaling_connected}a shows the scaling with size in case $BC$. We observe that $I(\rho_{BC})$ exhibits logarithmic scaling.  This corresponds with the expected scaling behavior for connected partitions, mirroring that of entanglement entropy in critical phases, which is well-established result in Conformal Field Theory \cite{Calabrese_2004, Vidal_2003}. However, what differs from entanglement entropy is the scaling of $I(\rho_{BC})$ with the bond dimension $\chi$, as illustrated in Fig. \ref{fig:mutualIscaling_connected}b. We focus on a single transition: the Haldane-Large D transition, and perform the scaling for $N\in\{16,28,40\}$. Notably, unlike what happens for $S_{N/2}$, the mutual information converges at a fixed value from above. What plays a role in this discrepancy are the partitions under examination. In our study, we analyze a four-partite system, unlike the bipartite case of entanglement entropy. Hence, the information extracted is not directly comparable to that obtained from the latter. In the case of $AC$, as depicted in Fig. \ref{fig:mutualIscaling_disconnected}a, we observe that the mutual information $I(\rho_{BC})$ remains constant with increasing size, consistently with previously known results \cite{Li_2019}. Fig. \ref{fig:mutualIscaling_disconnected}b shows the scaling with $\chi$ for the Haldane-LargeD transition and for the same three sizes as in the previous case. Even in this scenario, we observe a rapid saturation in the bond dimension.

As concerns long-range magic, the numeric results are challenging to interpret due to statistical error. In Fig. \ref{fig:LRMscaling_connected}a and Fig. \ref{fig:LRMscaling_disconnected}a, we display the scaling with the system size $N$ for case $BC$ and $AC$, respectively. Comparing with the behavior of mutual information, long-range magic appears to increase less rapidly for two connected partitions, while remaining constant for disconnected partitions. Turning to the scaling with the bond dimension $\chi$, depicted in Fig. \ref{fig:LRMscaling_connected}b, Fig. \ref{fig:LRMscaling_disconnected}b, we notice parallels with the behavior of mutual information. However, unlike for the full-state magic case, we are not able to predict a functional form for the correction in the bond dimension. Specifically, to estimate the $\chi$-dependence, in the previous section we focused on $\chi \geq 5$. Yet, at these values of $\chi$, it is difficult to discern the scaling, as the fluctuations we aim to fit are of the same order as the error bars with a sample size of $N_S = 10^6$. This is because our estimation of long-range magic comprises a sum of two terms, each associated with its own statistical error, and the $W$ term defined in Eq. \eqref{eq:estimator_w} is more challenging to estimate. As a consequence, we cannot predict the precise scaling of long-range magic in $\chi$, but we consistently observe rapid saturation.

\section{Conclusions}\label{sec:concl}
We have investigated the relation between magic and bond dimension in the context of ground states of spin-1 systems, for which the resource theory of magic in terms of stabilizer Renyi entropies is well under control. We considered both the full-state magic, a global property, as well as the mutual magic, that characterizes the magic that resides in the correlations between subsystems. We mostly show results at critical points, since inside phases, convergence of magic with the bond dimension is so quick that is hard to characterize.

For full-state magic, we provide extensive numerical evidence that the stabilizer Renyi entropies converge rapidly with a scaling of $1/\chi^2$, significantly faster than that of entanglement. Then, for mutual magic, we again observed a very mild scaling with respect to bond dimension, although we were not able to definitively determine the correct scaling form. These results lend credence to previous works that used small MPS bond dimension to study magic.

Additionally, a valuable byproduct of our investigation is the discovery that Pauli-Markov chains can efficiently estimate mutual information of disconnected subsystems, with the samples exhibiting small autocorrelations. This method - to the best of our knowledge - largely outperforms traditional methods utilizing exact TN contraction.

Overall, this work sheds light on the dependence of magic in MPS with the bond dimension; the latter is in turn directly linked to entanglement. Given the recent focus on the interplay between magic and entanglement, an interesting question is whether these results hold deeper significance, potentially hinting at a more fundamental connection between these two resource quantities. It would be intriguing to perform a similar analysis in the context of different classes of variational wave functions, such as projected entangled pair and tree tensor network states, where the bond dimension relation to entanglement is different than in MPSs. Finally, it could be interesting to investigate how some of the Pauli-based sampling methods might be suitably modified to measure magic and Renyi entropies experiments, relying on importance sampling.

\section*{Acknowledgments} 

We thank A. Hamma, T. Haug, G. Lami and L. Piroli for useful discussions and feedback on the manuscript, and M. C. Ba\~nuls, R. Fazio, G. Fux and X. Turkeshi for collaboration on related topics.
P.S.T. acknowledges support from the Simons Foundation through Award 284558FY19 to the ICTP. 
M.\,D. was partly supported by the MIUR Programme FARE (MEPH), by the EU-Flagship programme Pasquans2, by the PNRR MUR project PE0000023-NQSTI, and by the PRIN programme (project CoQuS).
M.C. was partially supported by the PRIN 2022 (2022R35ZBF) - PE2 - ``ManyQLowD''.
MD would like to thank the Institut Henri Poincaré (UAR 839 CNRS-Sorbonne Université) for their support. MD and ET were partly supported by QUANTERA DYNAMITE PCI2022-132919,

{\it Note added:} While completing this work, we became aware of a related study in the context of random matrix product states \cite{lami2024quantum} by Lami, Haug and De Nardis, that appeared on the same arxiv posting. 

\appendix
\section{Autocorrelations}
In this section we show a detailed analysis of the autocorrelation of the long-range magic. 

The normalized autocorrelation function of the stochastic process that generated the chain for $f$, denoted as $\rho_f(t)$, can be estimated for a finite set of $N_S$ samples, as 
\begin{equation}\label{eq:autocorr_func}
    \rho_f(t) = c_f(t)/c_f(0) \;,
\end{equation}
where 
\begin{equation}
    c_f(t) = \frac{1}{N_S - t} \sum_{n = 1}^{N_S - t} (f_n - \mu_f)(f_{n + t} - \mu_f)
\end{equation}
and 
\begin{equation}
    \mu_f = \frac{1}{N_S} \sum_{n = 1}^{N_S} f_n \;.
\end{equation}

We estimate the integrated autocorrelation time $\tau$ as 
\begin{equation}
    \tau = 1 + 2\sum_{t = 1}^M \rho_f(t)
\end{equation}
for some $M \ll N_S$. Extending the sum to $N_S$ is inconvenient because, for $t\gg\tau$,  $\rho_f(t)$ diminishes, leading to a situation where noise dominates over signal. Hence, introducing the cut-off $M$ helps to reduce the variance of the estimator $\tau$, at the cost of adding some bias. A good tradeoff between decreasing variance and introducing bias can be accomplished by choosing the smallest $M$ that satisfies $M \geq C \tau(M)$, tipically with $C\sim 5$. We estimated the autocorrelation times using the emcee library \cite{emcee} which employs the iterative procedure outlined in \cite{Sokal1997} to determine a suitable window size $M$.

Note that, for each computation of long-range magic, we make use of two Markov chains: one for the estimation of $I(\rho_{AB})$ defined in Eq. \eqref{eq:estimator_i2}, the other one for the estimation of $W(\rho_{AB})$, as in Eq. \eqref{eq:estimator_w}. The two autocorrelation times are thus not directly related.
\begin{figure}[t!]
    \centering
    \includegraphics[width=0.47\textwidth]{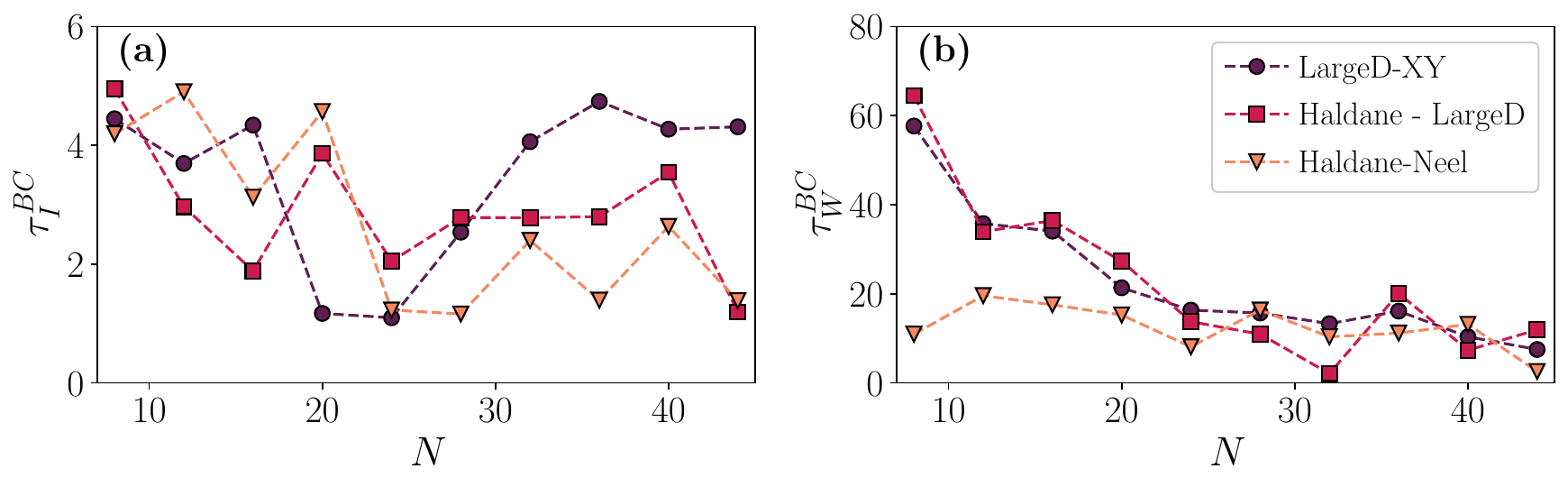}
    \caption{Integrated autocorrelation times scaling with size $N$ for connected partitions $B$, $C$ of length $L = N/4$ (see Fig. \ref{fig:partitions}). \textbf{(a)} Integrated autocorrelation time in the estimation of $I(\rho_{BC})$ (Eq. \eqref{eq:estimator_i2}). \textbf{(b)} Integrated autocorrelation time in the estimation of $W(\rho_{BC})$ (Eq. \eqref{eq:estimator_w}).}
    \label{fig:tau_connected} 
\end{figure}
\begin{figure}[t!]
    \centering
    \includegraphics[width=0.47\textwidth]{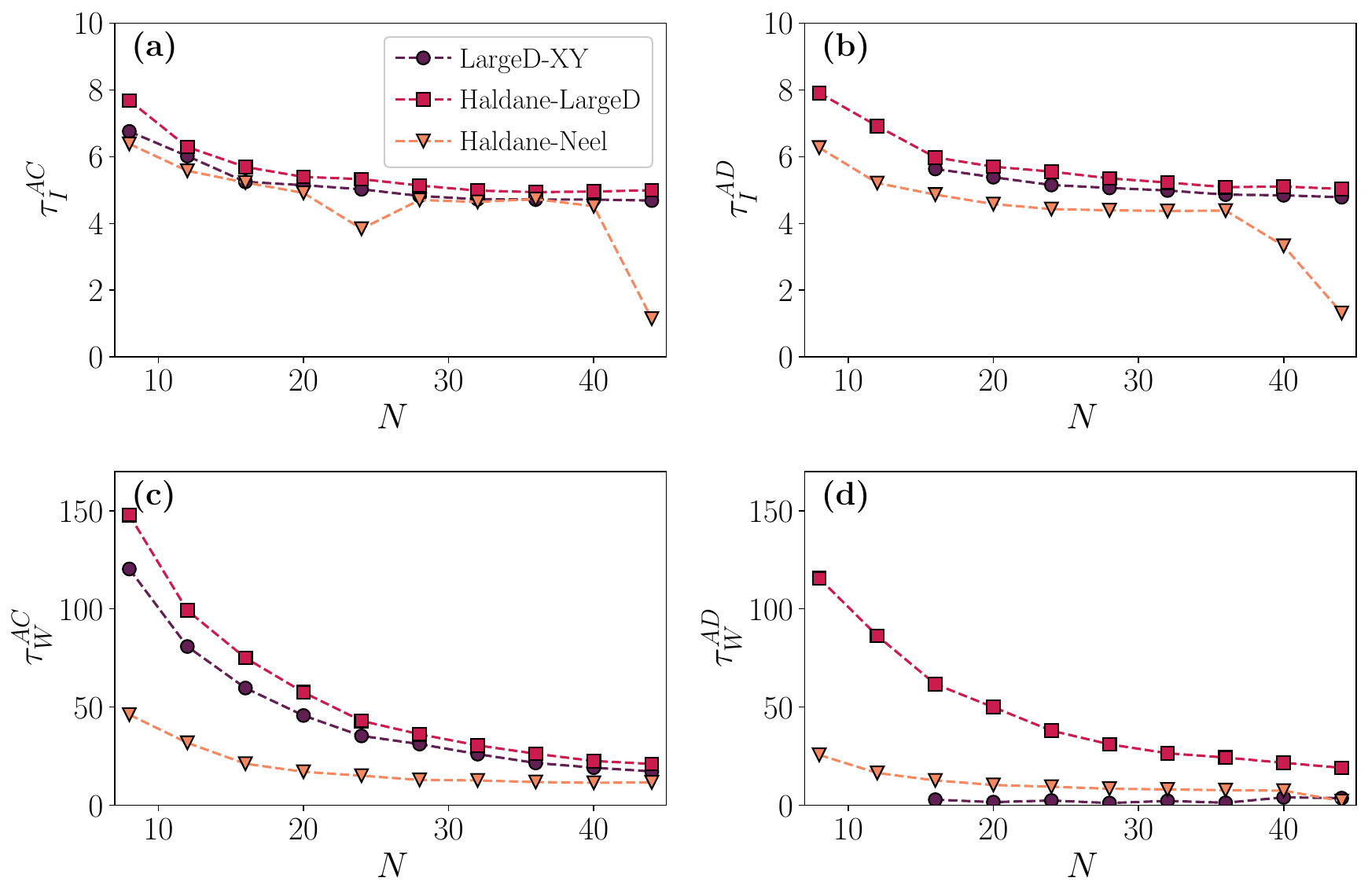}
    \caption{Integrated autocorrelation times scaling with size $N$ for disconnected partitions of length $L = N/4$. Panels \textbf{(a)} and \textbf{(c)} refer to the choice of partitions $A$ and $C$ in Fig.\ref{fig:partitions} (one in the boundary and one in the bulk, with distance $N/4$ between the two), while \textbf{(b)} and \textbf{(d)} refer to the choice $A$ and $D$ in Fig.\ref{fig:partitions} (both at the boundary, with distance $N/2$ between the two). Panels \textbf{(a)} and \textbf{(b)} show the integrated autocorrelation times in the estimation of  $I(\rho_{AC/AD})$ (Eq. \eqref{eq:estimator_i2}), while \textbf{c} and \textbf{d} show the integrated autocorrelation times in the estimation of $W(\rho_{AC/AD})$.}
    \label{fig:tau_disconnected}
\end{figure}

In Fig. \ref{fig:tau_connected}, we show the results for two connected partitions (as $B$, $C$ in Fig. \ref{fig:partitions}). We observe that increasing the size of the system $N$, $\tau_I^{BC}$ is always $\leq 5$, hence essentially constant. $\tau_W^{BC}$ is instead decreasing with size. 

Fig. \ref{fig:tau_disconnected} shows the results for disconnected partitions. We study two possible scenarios: both partitions at the boundary of the chain ($A$, $D$ of Fig. \ref{fig:partitions}) or one at the boundary and one in the bulk ($A$, $C$). We observe that again $\tau_I$ is approximately constant while $\tau_W$ is decreasing with size, for both cases. This decreasing trend can be understood in terms of our proposal of the candidate string: at each Markov step we update either a single site or two sites, that are randomly chosen. The updated sites can then fall in two different partitions. When the two partitions are disconnected, as the size of the system increases, so does the distance between the two, making it easier to generate non-correlated Pauli strings. Indeed, when the two partitions are more distant, namely in the $AD$ case, where they are separated by a distance $N/2$, the integrated autocorrelation time $\tau_W^{AD}$ is always smaller than the one corresponding to the other scenario $\tau_W^{AC}$, where the distance between the partitions is $N/4$.

\begin{figure}[b!]
    \centering
    \includegraphics[width=0.47\textwidth]{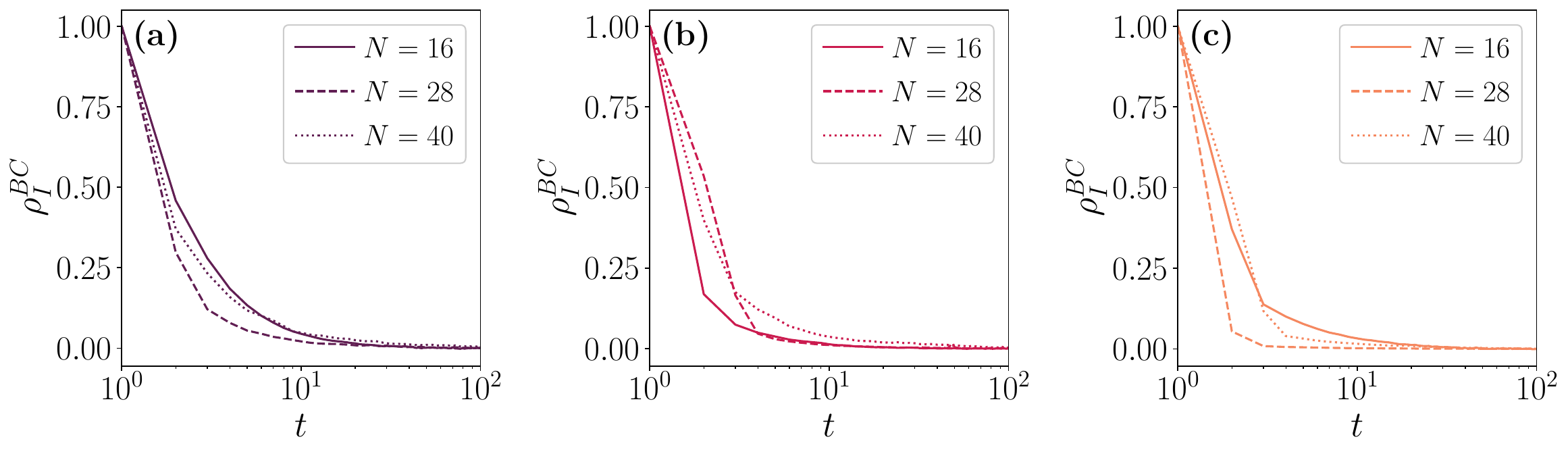}
    \caption{Normalized autocorrelation function of the mutual information $I$ in case $BC$, for different sizes. Panel \textbf{(a)}: Large D-XY transition (BKT at $J_z \sim -0.183$, $D \sim 0.5$). Panel \textbf{(b)}: Haldane-Large D transition (Gaussian at $J_z \sim 0.5$, $D \sim 0.635$). Panel (\textbf{c}): Haldane-Néel (Ising at $J_z \sim 2.93$, $D\sim 2.6$).}
    \label{fig:rho_I_connected}
\end{figure}

To further illustrate the origin of the above results, in Fig. \ref{fig:rho_I_connected}, we show the behavior of the normalized autocorrelation function of the mutual information $I$ for case $BC$. Figs.~\ref{fig:rho_I_connected}a, \ref{fig:rho_I_connected}b and \ref{fig:rho_I_connected}c correspond, respectively, to the Large D-XY, Haldane-Large D and Haldane-Néel transitions. Counterintuitively, we observe that despite the mutual information increasing with size (see Fig. \ref{fig:mutualIscaling_connected}), the autocorrelation of the Pauli-Markov chain remains constant.

These results demonstrate a remarkable unexpected efficiency of Pauli-Markov chains when dealing with long-range correlations: indeed, such Markov chains typically show a dynamical critical exponent between 0 and 1 for full state magic, while here, the exponent is actually negative - that is, sampling becomes simpler as volume increases, despite information not necessarily decaying. This feature makes Pauli-Markov chains potentially interesting to be used in experimental protocols.

\bibliography{bibliography}

\end{document}